\documentclass[NumberedRefs, preprint]{JASA-EL}
\usepackage{pagecolor}
\pagecolor{white}
\usepackage{amssymb}
\usepackage{natbib}
\usepackage{amsmath}
\usepackage{float}
\usepackage{cancel}
\usepackage{longtable}
\usepackage{comment}
\usepackage{graphicx}
\AtBeginDocument{\nolinenumbers}
\usepackage{etoolbox}
\makeatletter
\AtBeginDocument{%
  \@ifpackageloaded{lineno}{%
    \let\linenumbers\relax
    \let\nolinenumbers\relax
  }{}
}
\makeatother

\begin{document}

\title{Influence of string register locations on vibratos among violoncellists}
\author{Steven Hu}
\correspondingauthor
\email{stehu@umass.edu}
\affiliation{Department of Physics, Harvard University, Cambridge, MA 02138}
\author{Sophia H. Kim}
\email{sophiakim@college.harvard.edu}
\affiliation{Department of Physics, Harvard University, Cambridge, MA 02138}
\author{Helena H. Kim}
\email{hk3397@columbia.edu}
\affiliation{Department of Physics, Harvard University, Cambridge, MA 02138}
\author{Hugo Mackay}
\email{hugomackay@college.harvard.edu}
\affiliation{Department of Physics, Harvard University, Cambridge, MA 02138}
\author{Eric J. Heller}
\email{eheller@fas.harvard.edu}
\affiliation{Department of Physics, Harvard University, Cambridge, MA 02138}
\affiliation{Department of Chemistry and Chemical Biology, Harvard University, Cambridge, MA 02138}

\begin{abstract}
This study analyzes how vibrato changes with finger position along the cello string. Examining 94 excerpts, we found moving the finger toward the bridge strongly increases acoustic vibrato depth ($\rho=0.6902$, $p=1.408\cdot 10^{-14}$). However, the performer's physical finger amplitude simultaneously decreases ($\rho=-0.6391$, $p=4.172\cdot 10^{-12}$). This shows players reduce finger motion in higher positions, but not enough to counteract the greater pitch deviation there, revealing both the presence and limits of compensatory vibrato behavior.

\end{abstract}
\maketitle
\section{Introduction}
Vibrato is a critical component of musical expression across almost all non-percussive instruments, and the vibrato of string instruments has received particular analytical attention due to its interaction with the instruments’ natural resonances and the visible, tangible physical mechanism by which it is produced. Broadly speaking, there are two factors that define a vibrato: frequency and depth. The depth of vibrato is of particular interest to us, as its physical and perceptual implications vary widely depending on context, including the instrument, register, and finger position \citep{Mellody2000, Fritz2010}. 

Thus, the study of vibrato depth is notably challenging and multifaceted, and various efforts have been made to understand the depth of vibrato since the inception of the field of Musical Acoustics \citep{Seashore1938}. The expected range of string player vibratos is reported to typically range from 0.2 to 0.35 semitones in depth \citep{Meyer1992}, and earlier foundational work by Shonle and Horan (1980) systematically examined violin vibrato rates and extents, establishing a baseline for later acoustical and perceptual analyses \cite{Shonle1980}. Brown and Vaughn (1996) demonstrated that listeners tend to perceive the ``center" of a vibrato tone as lying closer to the mean of the pitch oscillation rather than either of its extremes \cite{Brown1996}. As the understanding of vibrato continued to advance, more studies started acknowledging how depth perception is described by the oscillations in power and spectral features. For instance, Verfaille et al. (2005) proposed models that treated vibrato not only as a modulation of fundamental frequency, but also of amplitude and spectral envelope, and demonstrated that these non-pitch modulations are perceptually significant \cite{Verfaille2005}. 
    
The vibrato depth in the context of the performer is of particular interest, as it features a scenario in which the instrumentalist is a listener who actively adjusts their performance based on what they perceive. Some studies have looked into how vibrato performance varies with experience. A 2012 study compared high school and university string players and found consistent differences in vibrato width and rate, demonstrating that both performance practice and listener experience shape how vibrato depth is perceived\cite{Geringer2012}.

To focus our investigation on vibrato performance, we sought to quantitatively investigate how the depth of vibrato relates to the pitch that is played. It is well known that the pitch of a note performed on a stringed instrument is inversely proportional to the length of the string between the finger and the bridge. It then follows that a vibrato performed with the same physical depth closer to the bridge as a vibrato farther away from the bridge would have a greater oscillation in pitch. If a cellist were to vibrate 1/10th of a semitone around E4 (1/3 of the way down the string, frequency = 330 Hz) on the A string, the same amount of physical motion around E7 (11/12 of the way down the string, frequency = 2640 Hz) would produce a vibrato of over 5 and half semitones in depth.

MacLeod’s 2008 study on high school and university-aged musicians has suggested the existence and relevance of this phenomenon in performance, demonstrating that vibratos are considerably wider in pitch in higher string registers\citep{MacLeod2008}. Additional review work in 2010 has further summarized the range of empirical findings on vibrato, widths, pitch center, and continuity across studies and notes explicitly remaining gaps in understanding how acoustical vibrato depth scales continuously along the fingerboard across all finger positions \cite{Geringer2010}. A 2024 study which analyzed the vibratos of university double bassists found a similar increase in width and faster vibrato rates as they performed in progressively higher registers \citep{Mick}.

While the results of these studies are consistent with physical intuition, they also have limitations that call for further research. For example, the studies by MacLeod and Mick provide limited detail about the precise frequencies used \citep{MacLeod2008, Mick}. While MacLeod’s 2008 study distinguishes between “high pitch” and “low pitch,” and Mick’s 2024 study refers to string positions (first through fourth), these categories show general trends, but a more precise, quantitative analysis is needed to understand the effects in detail.

Furthermore, while the results of these studies are intuitive, there is still one potentially confounding factor to be addressed: with the same finger movement in high and low registers, acoustical vibrato depth would drastically increase with register (see Fig. 4). First, the question that arises is how players compensate for this. Do they partially or completely counteract the effect? As one increases in register, is there any compensatory action done to restrict or limit the acoustical vibrato depth? After all, the performer is both the performer and the listener, actively responding to their own vibrato. The performer may likely be aware that as they get closer to the bridge, the vibrato widens, and they may purposely 'compensate' their physical finger movement. In fact, some studies support this hypothesis, suggesting a decreasing relationship between physical depth and frequency of vibrato, suggesting that a restriction in physical motion allows for a greater vibrato rate at the same finger speed\citep{TimmersDesain2000} (see Supplementary Material).

This study examines the correlation of the register of note on vibrato performance in a more rigorous light by introducing a physical depth parameter and consequently allowing an analysis of a continuous data set (i.e. spanning the full length of the fingerboard).

Because changes in vibrato physical depth may be confounded by limitations in finger control, we restricted our dataset to recordings by professional-class performers, in contrast to prior studies that primarily feature student musicians. Ultimately, this study is expected to contribute to a more complete understanding of the physical execution of vibrato in stringed instruments and the resulting influence on the way it is perceived. We also hope that our methods of analysis may be extrapolated to vibrato across other musical instruments and genres.
\section{Method}
We extracted a total of 94 vibrato files, 40 from the TinySOL soundbank and 54 from commercial recordings of world-class performers. We first examined files from the TinySOL soundbank as they were recorded in a controlled setting with similar equipment and environmental conditions. We then similarly studied 54 excerpts from commercial recordings, which were done in live performance venues, and were subject to more variation with different acoustic conditions and recording settings. All 54 performance files were selected from concerto recordings, indicating the presence of orchestral accompaniment. 

Our analysis primarily used Librosa (v. 0.11.0), a multi-feature Python package which includes a YIN function that utilizes the autocorrelation function to provide an estimate of a pitch at a certain time $t$\citep{CheveigneKawahara2002}. We used a simple YIN function with a restricted frequency search range ($\approx$ intended center pitch $\pm$ 200 cents, to be adjusted). Then, we plotted the value of the YIN function with respect to time to track the pitch oscillations in a performed vibrato.

To account for variations in string length across different cellos, we expressed the physical position and depth of vibrato in terms of a ratio. The physical position of the finger on the string can be expressed in terms of the ratio between the frequency of the string and the frequency of the note at a certain time $t,$ or 
\begin{equation}
x=1-\frac{f_{\text{s}}}{f_{\text{t}}},
\end{equation}
where $x$ is a unitless ratio, $f_{\text{s}}$ is the frequency of the string and $f_{\text{t}}$ is the frequency of the tone. Hence, a higher $x$ value indicates a greater proximity to the bridge of the instrument, while a lower $x$ indicates a greater proximity to the nut. For example, $x=0.5$ on an average full-sized cello (with string of length $\sim$70cm\citep{ThomastikInfeld, Pirastro}) corresponds to 35cm in length from the nut. Furthermore, a physical vibrato depth of $D=0.01$ would correspond to $0.7\text{cm}$ above and below the physical center. 

\subsection{Audio file selection}
All vibrato files used in this study were selected from one of two sources: the TinySOL sound bank and commercially recorded cello performances \citep{TinySOL}. The vibrato files extracted from these recordings were selected for analysis if they satisfied the following criteria: (1) the excerpt contains a \textit{singular} note at least a second in length, and (2) the soloist is the primary voice. 

To test criterion (2), the YIN vs. Time plot was examined visually for irregularities, and files were deemed satisfactory if the plot appeared reasonably sinusoidal. Audacity\citep{Audacity2025} was used to trim excerpts from recordings as well as exclude the attack and release of notes.

\subsection{Parameter calculations}
While acoustical vibrato depth is often defined as the difference between the maximum and minimum frequencies of a tone, we chose to adopt a more rigorous measure to address inconsistencies and pitch drift throughout the course of a note (i.e. the tonal center fluctuating throughout the course of the note). To reduce microfluctuations in the YIN contour, a light low-pass filter ($\text{savgol\_window\_max}=101$, $\text{savgol\_polyorder}=3$) was applied to the YIN function plot. This method is commonly used in bioacoustics and pitch processing applications to reduce noise in oscillatory signals without attenuating the amplitude of relevant modulations. Next, peaks and troughs of the smoothed plot were found and indexed using SciPy's (v. 1.16.0) find\_peaks function. 

We found midpoints between consecutive peaks and troughs, and connected them by straight-line interpolation to form a ``trend line". Finally, distances between peaks, troughs, and the trend-line were averaged to provide an accurate estimate of the average acoustical vibrato depth ($d$). An example of this process may be seen below in Figure 1:

\begin{figure}[H]
\centerline{\includegraphics[width=6in]{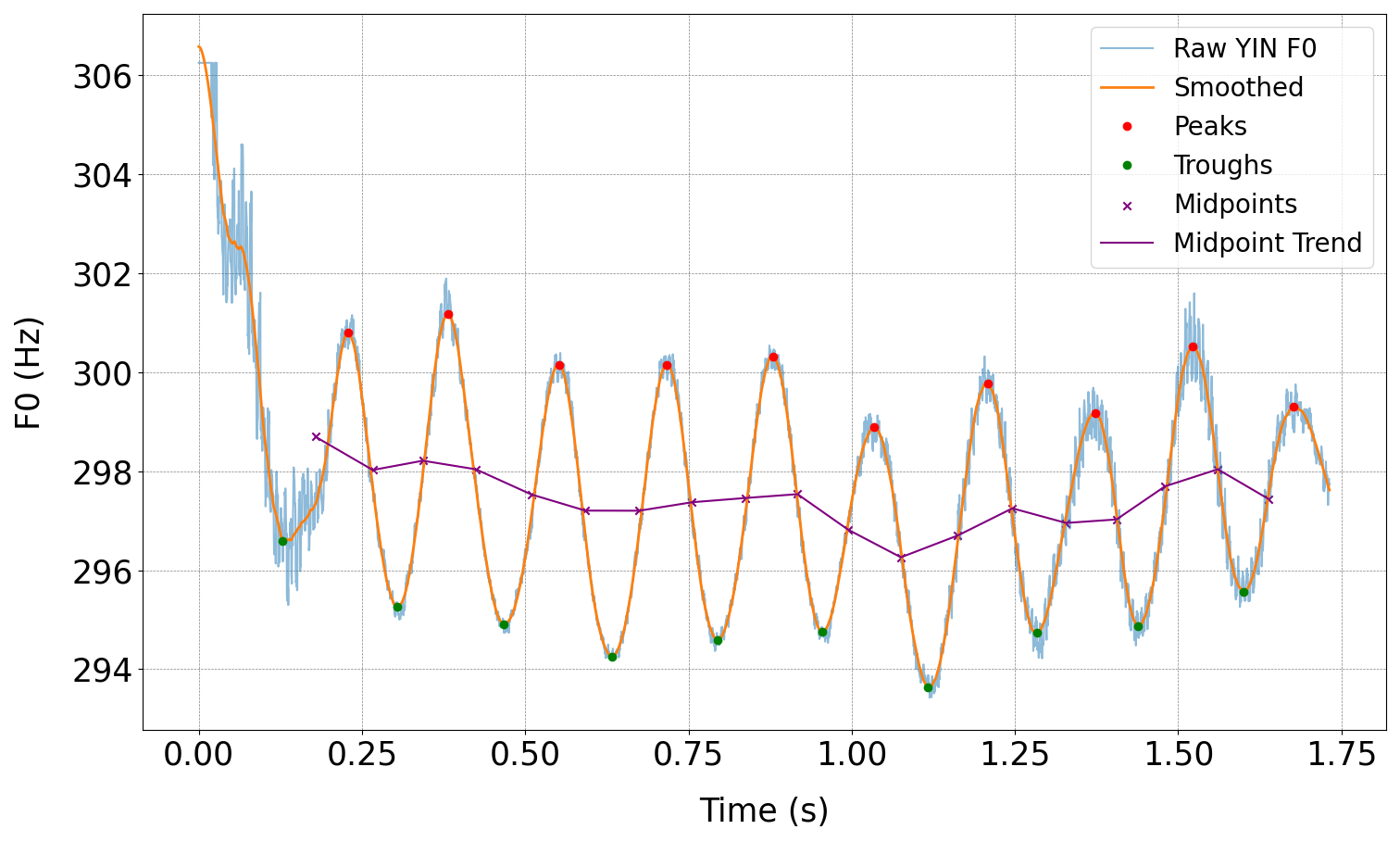}}
\caption{\label{fig:FIG1}\textbf{Pitch contour extracted via YIN from pfD4.wav}, low-pass filtered, with detected peaks and troughs indicated. Acoustical vibrato depth is quantified as the average deviation from the midpoint trend line—rather than the conventional peak-to-peak pitch range—to reduce volatility in the measurement.}
\raggedright
\end{figure}
This was followed by a calculation of the physical depth ($D$) of  finger oscillation during the vibrato as a proportion of the total string length as follows:
\begin{equation*}
2\cdot d = f_s \cdot \left(\frac{1}{(1-x_c)-D}-\frac{1}{(1-x_c)+D}\right)
\end{equation*}
\begin{equation*}
d\cdot D^2 +f_s\cdot D-d\cdot (1-x_c)^2=0
\end{equation*}
\begin{equation*}
D=\frac{-f_s+\sqrt{f_s^2+4d^2(1-x_c)^2}}{2d}
\end{equation*}
Combining this with equation (1) yields
\begin{equation}
D=\frac{f_s}{2d}\left(\sqrt{1+\frac{4d^2}{f_c^2}}-1\right)m
\end{equation}
where $f_c$ denotes the tonal center of the vibrato note, and $x_c$ the corresponding physical center (according to equation (1)). Like $x$ the physical position of the finger on the string, $D$ is a proportion of the total string length and is thus also unitless. 

We examined a total of six quantities for all selected files: (1) the average depth of pitch variation in cents, (2) the average depth of pitch variation in Hz, (3) the tonal center of vibrato (Hz), (4) the physical depth of vibrato as a proportion of the total string length, (5) the physical center of vibrato, as a proportion of the total string length, and (6) the vibrato rate (Hz), which refers to the rate at which a performer vibrates, expressed as the multiplicative inverse of the average x-axis distances between consecutive peaks in the YIN vs. Time plot. Of the six quantities, we examined average depth of pitch variation in cents, physical depth of vibrato as a proportion of the total string length, physical center of the vibrato, and vibrato rate in greater detail by examining scatter plots to investigate relationships.
\section{Results}

Figures 2 and 3 plot the parameters extracted from commercial recordings (blue) and TinySOL soundbank (red). The x-axis for both of these plots indicates the physical center of vibrato as defined in the introduction ($x_c$). Polynomial regressions were completed up to the second order.

\begin{figure}[H]
\centerline{\includegraphics[width=6in]{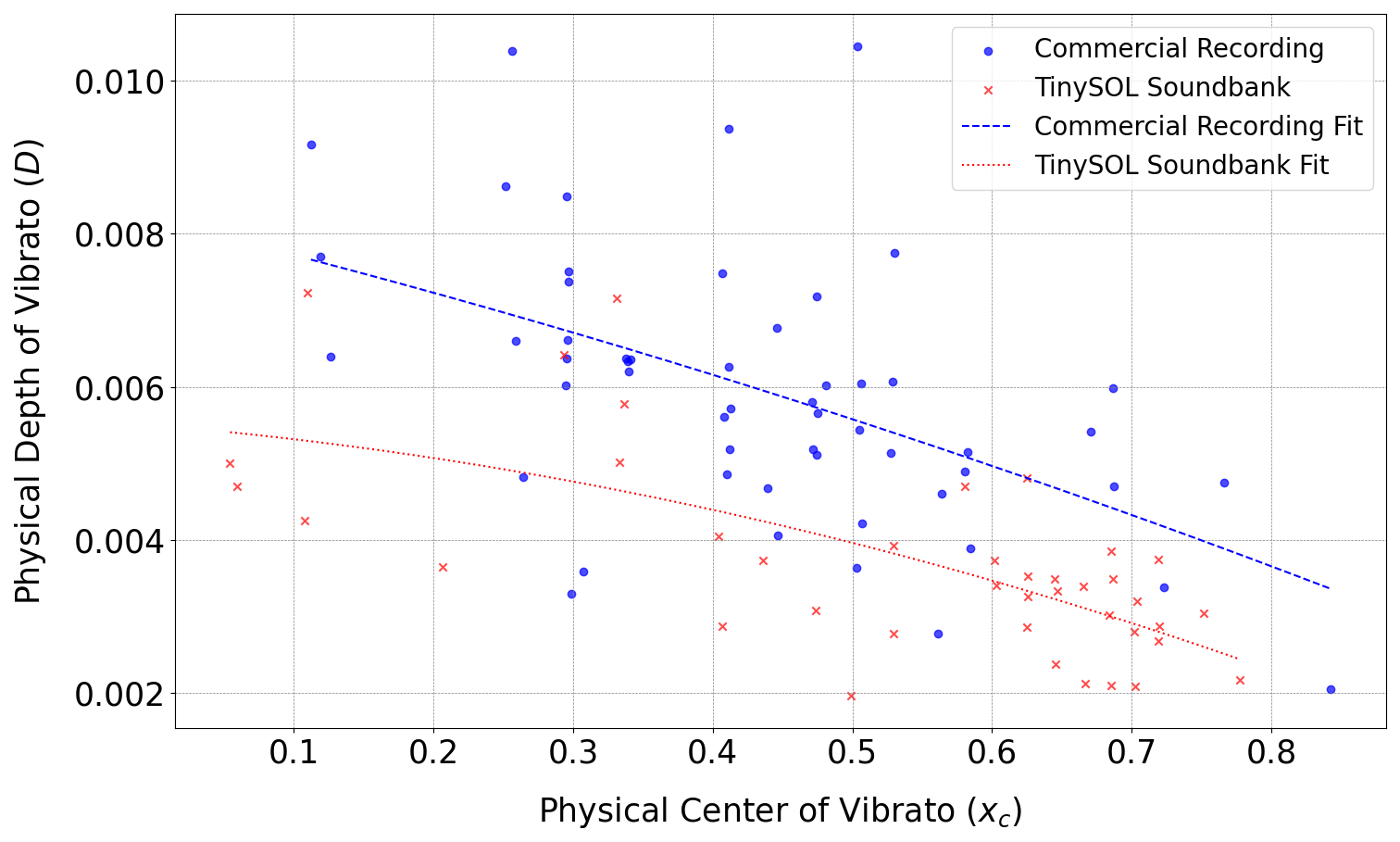}}
\caption{\label{fig:FIG2}\textbf{The physical depth of vibrato decreases as the vibrato center moves toward the bridge.} A scatterplot of the physical center of vibrato (as a proportion of string length) and the physical depth of vibrato (also as a proportion of string length) across all 94 samples exhibits a general decreasing trend. Additionally, the commercial recordings (blue) show a notably greater spread of physical depth and an overall greater physical depth across all registers.}
\raggedright
\end{figure}

Figure 2 is a scatter plot of physical vibrato depth as a proportion of total string length against the physical center of the vibrato, also expressed as a proportion of the string length from nut to bridge. A modest downward trend appears between the physical depth of vibrato finger motion and the physical center of the vibrato on the string. The data show no notable clusters on either axes. 

For TinySOL files (indicated in red), there is a moderate negative correlation between the physical depth of vibrato finger motion and the physical center of the vibrato on the string ($R=-0.66,$ $R^2=0.44$). A decreasing quadratic polynomial ($y=-0.0031(x - 0.25)^2 + 0.0057$) also provides a moderate fit ($R^2=0.45$). For commercial files (blue), there is a moderate negative correlation between the variables ($R=-0.52,$ $R^2=0.27$). A decreasing quadratic polynomial ($y=-0.0015(x +1.5)^2 + 0.012$) also fits moderately ($R^2=0.27$). For the combined set (black), there is a moderately weak negative correlation between the variables ($R=-0.58,$ $R^2=0.34$). A decreasing quadratic polynomial ($y=-0.0079(x - 0.054)^2 + 0.0066$) also fits moderately ($R^2=0.38$).

\begin{figure}[H]
\centerline{\includegraphics[width=6in]{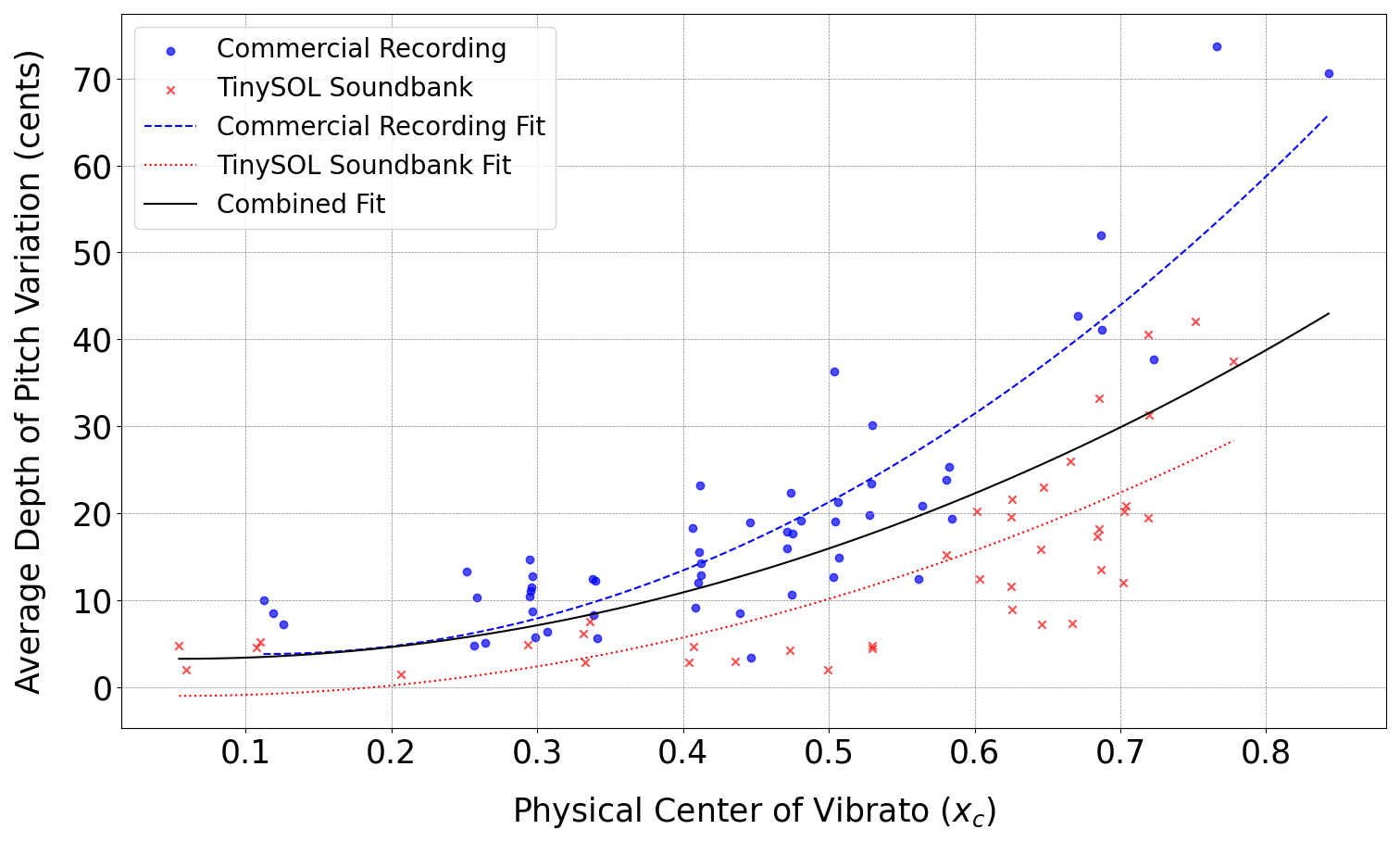}}
\caption{\label{fig:FIG3}\textbf{Pitch deviation increases markedly as the vibrato is executed closer to the bridge.} A scatterplot of the physical center of vibrato (as a proportion of string length) and pitch variation (in cents) shows a strong increasing trend. Here, the data cluster a lot closer to the fit curves, and while the data from the  commercial recordings still appear to have a greater spread than the data from the TinySOL recordings, it is much less pronounced.}
\raggedright
\end{figure}
Figure 3 plots the average depth of pitch variation in cents against the same physical center values. Here, a positive correlation is more pronounced: as the vibrato center approaches the bridge of the instrument, the pitch variation in cents increases. The data do not show notable clustering on either axes.

For TinySOL files, there is a moderate positive correlation between the physical center of vibrato and pitch variation (in cents) ($R=0.70,$ $R^2=0.48$). An increasing quadratic polynomial ($y=56(x - 0.054)^2 -0.953$) provides a moderately strong fit ($R^2=0.62$). For commercial files, there is a moderately strong positive correlation between the variables ($R=0.81,$ $R^2=0.65$). An increasing quadratic polynomial ($y=116(x - 0.112)^2 + 3.85$) provides a moderately strong fit ($R^2=0.80$). For the combined set, there is a moderate positive correlation between the variables ($R=0.65,$ $R^2=0.42$). An increasing quadratic polynomial ($y=63.8(x - 0.054)^2 + 3.31$) also fits moderately ($R^2=0.51$).

\section{Discussion \& Conclusion}
This study examined how cellists adjust their vibrato to account for variable effective string lengths of the notes they play. We analyzed a total of 94 samples from commercial recordings and the TinySOL sound bank to determine how performers compensate for this effect. Our analysis reveals two key findings concerning the depth of vibrato across different string register locations.

First, we observed a modest, positive linear correlation between the physical center of the vibrato on the string and the physical depth of the finger's oscillation (Fig. 2). A Spearman's test for monotonicity for these 3 data groups yields $\rho = -0.5560,$ and $p=1.271\cdot 10^{-5}$ for the commercial recordings, $\rho = -0.6366,$ and $p=1.007\cdot 10^{-5}$ for the TinySOL recordings, and $\rho = -0.6391,$ and $p=4.172\cdot 10^{-12}$ for the combined set.

This indicates that performers do make a physical adjustment, tending to reduce finger motion slightly for notes played closer to the bridge, suggesting a conscious or subconscious attempt to compensate for the string's physics to achieve a more consistent vibrato sound. However, acoustical vibrato depth also varies stylistically, and the string instrument technique may involve wider or narrow vibratos for expression and varying musical context. The variations due to such may explain our lack of a higher $R^2$ value in our regressions. 

However, our second finding demonstrates that this compensation is incomplete. Despite the decrease in physical motion, we found a strong positive correlation between the vibrato's physical center and the resulting average depth of pitch variation in cents (Fig 3.). These findings are consistent with prior studies \citep{MacLeod2008, Mick}. Furthermore, a Spearman's test for monotonicity for these 3 data groups yields $\rho = 0.7973,$ and $p=5.330\cdot 10^{-13}$ for the commercial recordings, $\rho = 0.8512,$ and $p=8.539\cdot 10^{-11}$ for the TinySOL recordings, and $\rho = 0.6902,$ and $p=1.408\cdot 10^{-14}$ for the combined set.

Even with narrowing vibrato at higher registers, the performer's physical adjustment is not sufficient to counteract the non-linear acoustic effect of shortening the string. As mentioned previously, this effect may in part be intentional and related to the aesthetics of performance. Musicians often view higher registers as more expressive and may desire to embellish a wider vibrato. 

The following figure summarizes these two findings:

\begin{figure}[H]
\centerline{\includegraphics[width=6in]{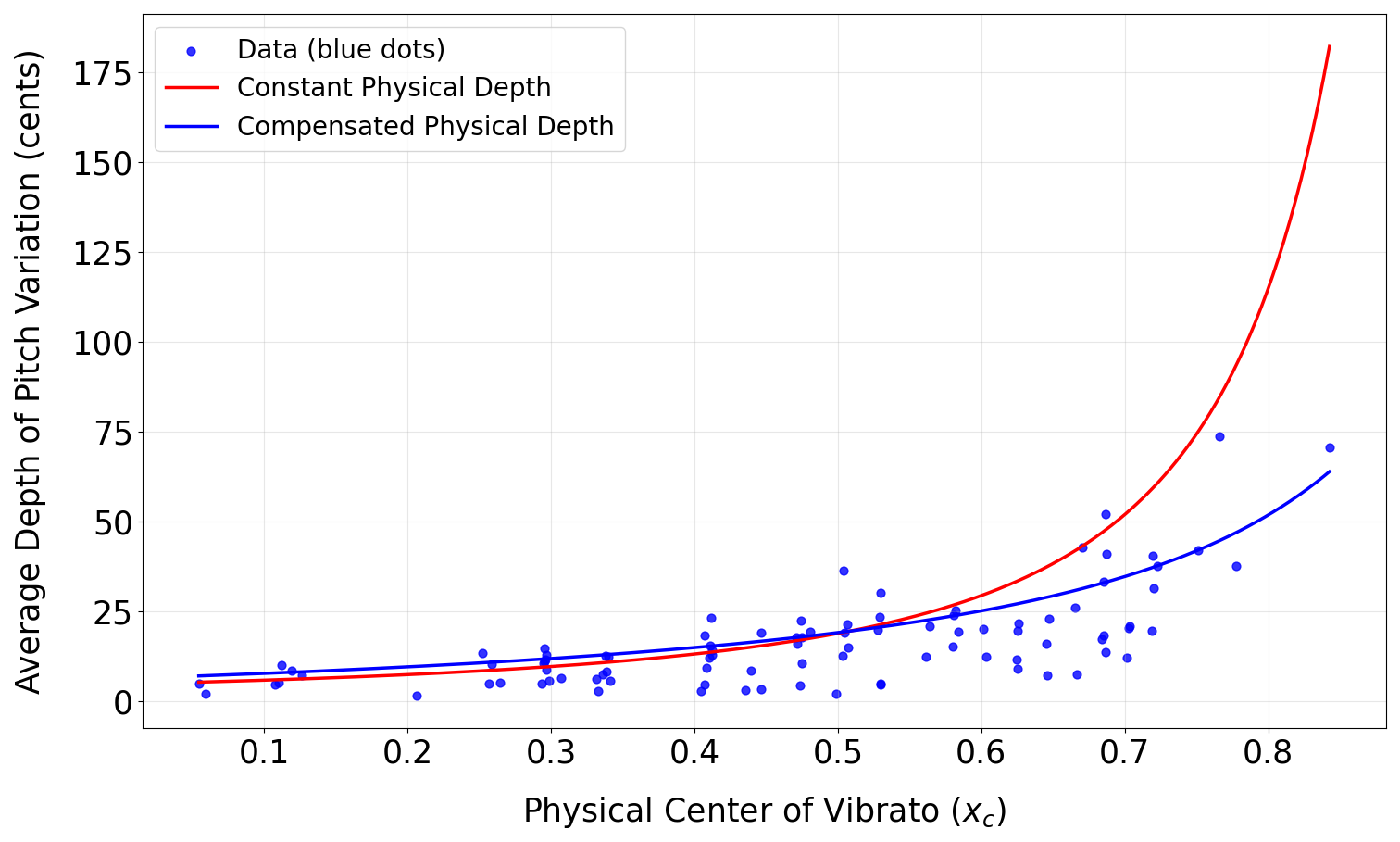}}
\caption{\label{fig:FIG4}\textbf{Pitch variation (cents) vs. physical vibrato center, shown with and without compensation.} Results from Figure 2 were combined with equation (2) to visualize the extent of the instrumentalist’s compensation. These two plots cross at around $x_c=0.5,$ indicating that the player’s compensation is effectively neutral when the vibrato is centered at the midpoint of the string.}
\raggedright
\end{figure}

The red line is the model of the pitch variation in cents under the hypothesis that the player maintains a constant width of finger motion regardless of the center of the vibrato. We chose this value to be the mean physical depth of our combined data set ($D=0.00497$). This yields an $R^2 = -1.06$. We can observe the line beginning slightly \textit{below} the data as the physical center of vibrato is closer to the nut of the instrument, and rapidly rising as the physical center of vibrato approaches the bridge.

The blue line is the model of the pitch variation in cents with a reduction of physical depth determined by the quadratic regression in Fig. 2 ($D=-0.0079(x_c-0.054)^2 + 0.0066$). Notice that this line aligns more closely with the data and crosses the uncompensated model at $(0.51, 19.54)$. This fit yields an $R^2=0.47.$

Although the polynomial fit in Fig. 3 reports a slightly higher $R^2$ value ($R^2=0.51$), this is likely an artifact of overfitting, given that it is not grounded in any physical model.

A likely contributor to the incomplete compensation of acoustical vibrato depth is biomechanical constraint. The finite width of the fingertip limits the extent to which physical vibrato amplitude can be reduced while maintaining a rocking motion; beyond this limit, further compensation would require translational sliding along the string. Consequently, full compensation is mechanically feasible only over a restricted range of pitch centers. Vocal vibrato may be subject to fewer analogous constraints, though the limited pitch range of the human voice restricts the extent to which this comparison can be explored.

More broadly, these conclusions contribute to our general understanding of vibrato. The general belief that string player vibratos range from $0.2$ to $0.35$ (or $0.1$ to $0.175$, as we calculated depth as half of the full peak-to-peak depth) semitones in depth is an oversimplification\citep{Meyer1992}. According to our combined quadratic regression in Fig. 3, this range corresponds to vibratos centered at normalized string positions between 0.377 and 0.526 (measured as fractions of the total string length), a commonly used—though not exhaustive—register in string performance.

In future studies, we hope to extrapolate these results to other stringed instruments, as well as non-stringed instruments. We expect similar results for violinists and violists, as the physical mechanisms to produce vibratos are extremely similar. For vocal and wind instruments, there does not involve a comparable confounding variable analogous to the physical depth of the finger variation on a string. Furthermore, more work remains to be done regarding the influence of nonpitch parameters in the perceived depth of performed vibrato, as variations in power (tremolo) and spectral features (spectral flux, spectral centroid, etc.) are extremely prominent in vibratos performed across many musical instruments.
\section*{AUTHOR DECLARATIONS}\vspace{-7pt}
\subsection*{Conflict of Interest}
The authors declare no conflict of interest.
\section*{DATA AVAILABILITY}
While commercial recordings cannot be shared due to copyright, all code and processed data supporting this study are openly available at \url{https://doi.org/10.5281/zenodo.17712642}. The commercial recordings analyzed in this work were accessed through licensed streaming services and are not redistributed according to copyright restrictions, while the TinySOL files can be found at \url{https://zenodo.org/records/3685331}.

\bibliographystyle{unsrt}
\bibliography{Vibrato}
\pagebreak
\section{Supplementary Material}
As mentioned in the main text, the tendencies of the individual performers can significantly influence the depth of vibrato beyond any easily traceable parameters. Fig S1, seen below, displays the physical center of vibrato vs pitch variation for commercial recordings, while highlighting the data and the quadratic fits for individual players:
\renewcommand{\thefigure}{S\arabic{figure}}
\setcounter{figure}{0}
\begin{figure}[H]
\centerline{\includegraphics[width=6in]{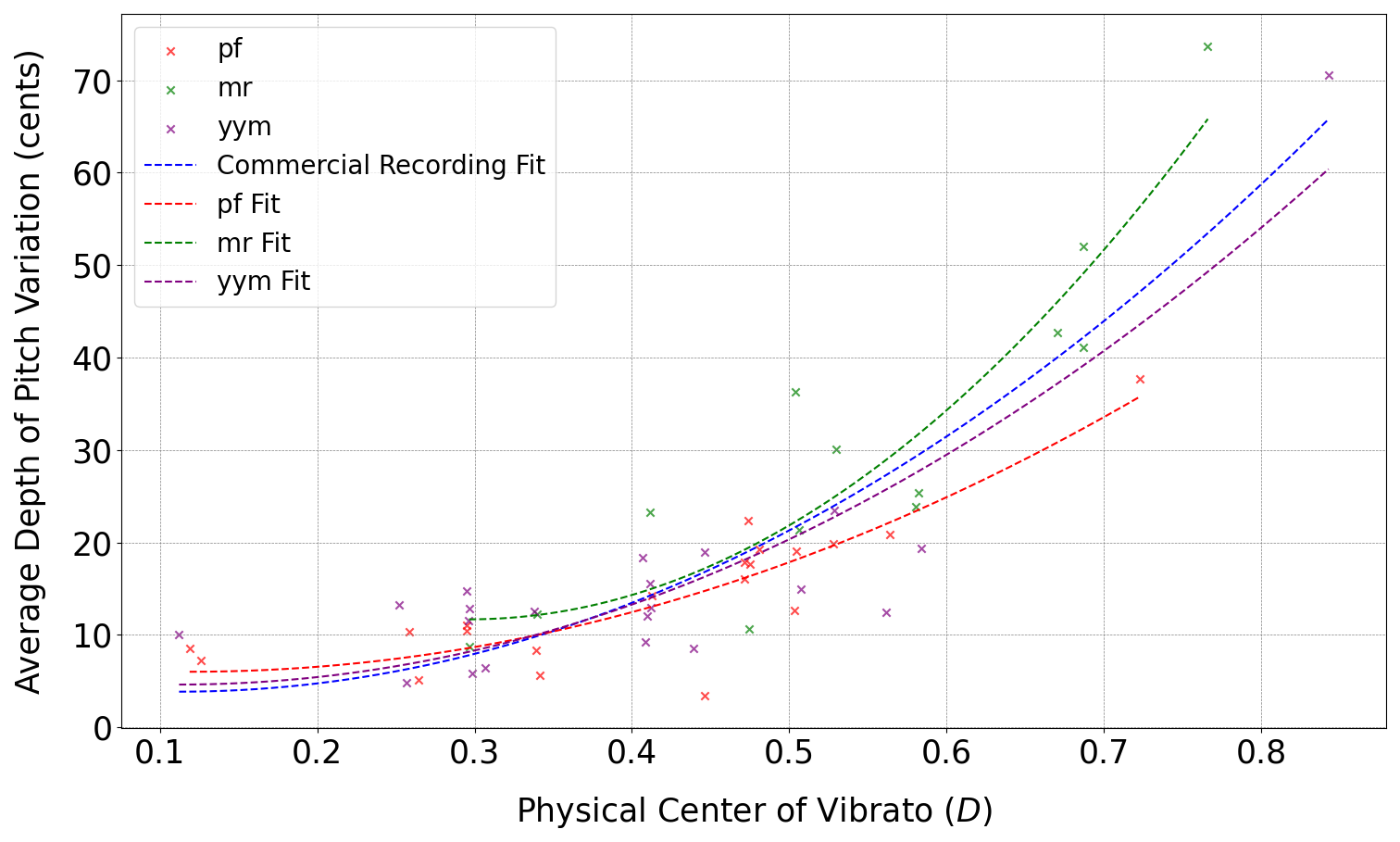}}
\caption{Relationship between physical center of vibrato and pitch variation (in cents) for commercial recordings, with data from individual players (Pierre Fournier, Mstislav Rostropovich, Yo-Yo Ma) and their respective fits highlighted.}
\raggedright
\end{figure}
The above diagram shows crossing in between quadratic fits for vibrato files from Fournier and Ma, while the quadratic fit for vibrato files from Rostropovich lies completely above the other two, as well as the combined fit. This provides reason to believe that Rostropovich vibrates systematically wider at all string registers, however, the limited sample sizes for each performer (20, 21, and 13, respectively) may account for this. We suggest that future studies may utilize large sample sizes to more conclusively determine if certain performers systematically prefer to vibrate deeper than others.
\begin{figure}[H]
\centerline{\includegraphics[width=6in]{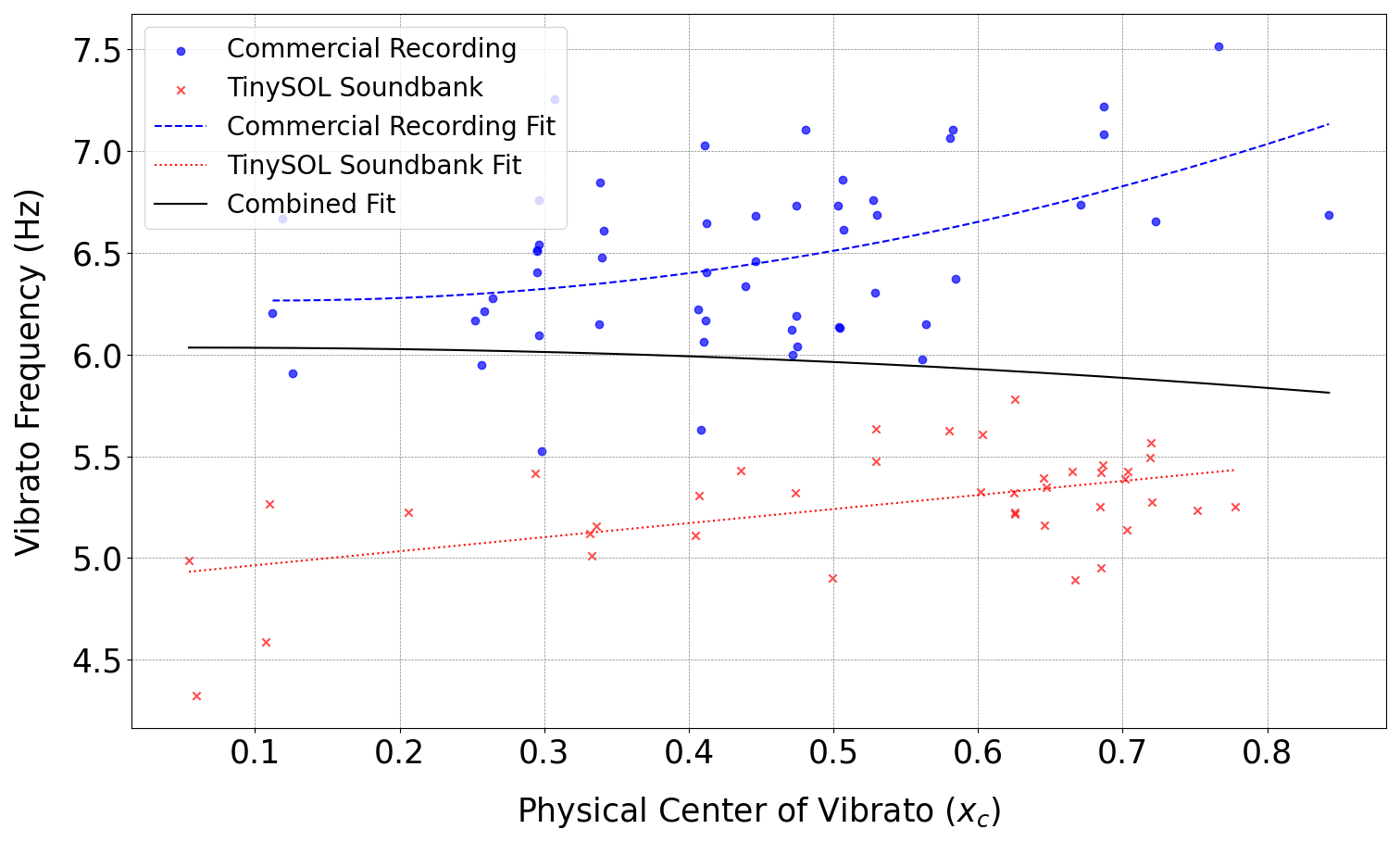}}
\caption{Relationship between physical center of vibrato and vibrato rate. Vibrato rate increases modestly as the performed note approaches the bridge of the instrument. }
\raggedright
\end{figure}
Figure S2 plots the vibrato rate against the physical center of the vibrato. Here we find a modest positive correlation for the commercial recordings and TinySOL soundbank, but no correlation for the combined set.

For TinySOL files, there is a moderately weak positive correlation between physical center of vibrato and vibrato rate ($R=0.51,$ $R^2=0.26$). An increasing quadratic polynomial ($y=0.00050(x - 685)^2 -231$) also fits moderately weakly ($R^2=0.26$). For commercial files, a linear regression ($y=1.10x + 6.00$) has a weak positive correlation with the data ($R=0.41,$ $R^2=0.17$). An increasing quadratic polynomial ($y=1.63(x - 0.11)^2+6.27$) fits weakly ($R^2=0.19$). 

The combined dataset, however, exhibits a weak negative trend ($R^2 < 0.01$). This is likely driven by the substantial difference in vibrato rate between the commercial recordings ($\bar{x}=6.475,\ s=0.4192$) and the TinySOL recordings ($\bar{x}=5.261,\ s=0.2770$), which effectively prevents treating the two datasets as a single population, even though both span the full length of the fingerboard.

A Spearman's test for monotonicity for these 2 data groups yields $\rho = 0.3739,$ and $p=0.005352$ for commercial recordings, and $\rho = 0.3066,$ and $p=0.05435$ for TinySOL recordings.
\begin{figure}[H]
\centerline{\includegraphics[width=6in]{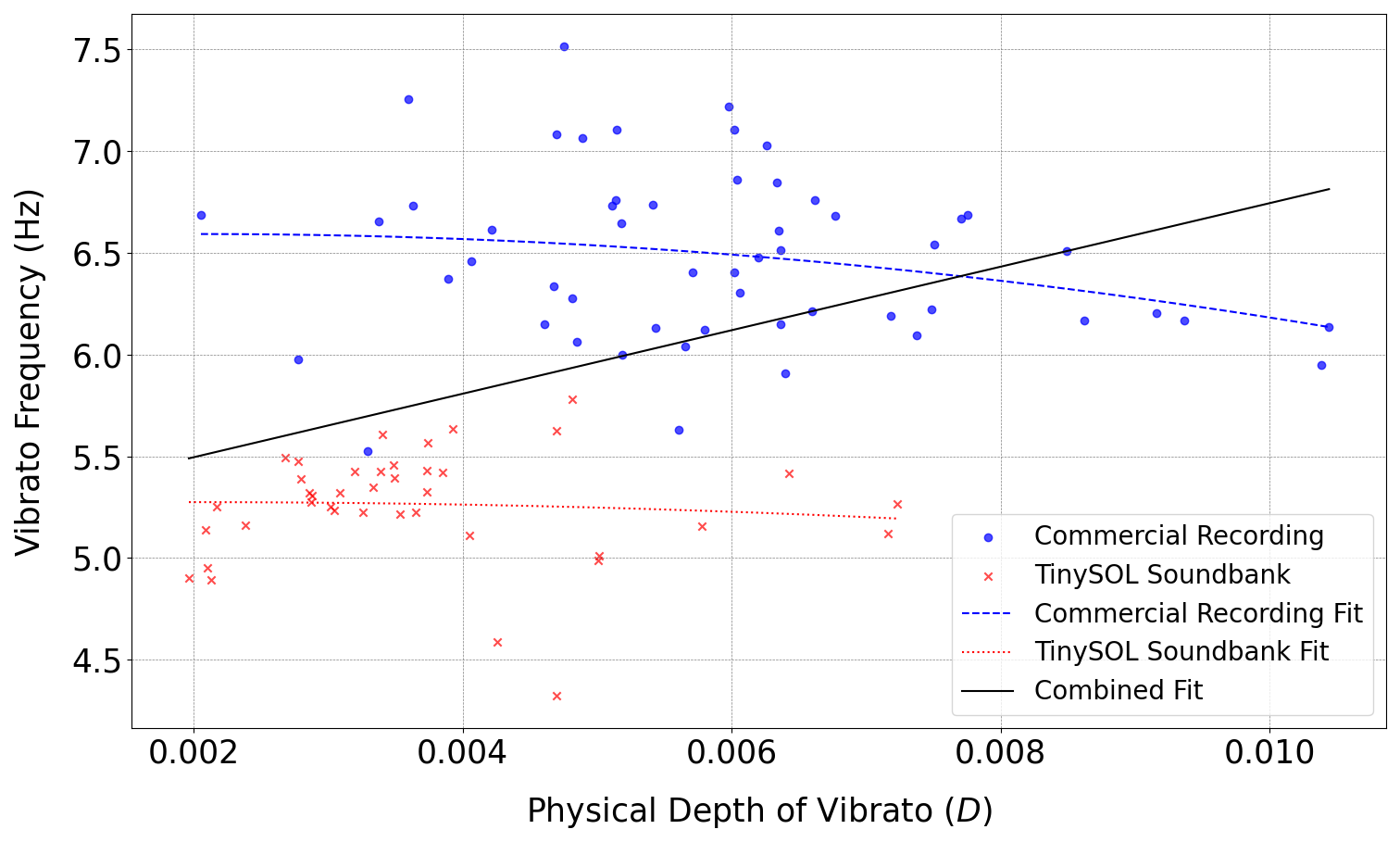}}
\caption{\label{fig:FIG5}Relationship between physical depth of vibrato and vibrato rate. Vibrato rate decreases modestly as the performed note is produced with smaller physical depth.}
\raggedright
\end{figure}
Figure S3 plots the vibrato rate against the physical depth of the vibrato. Here we find no correlation for TinySOL soundbank or commercial recordings, while a weak positive correlation emerges when combining the two sets of recordings.

For TinySOL files, there is an extremely weak negative correlation between the physical depth of vibrato and the vibrato rate. A decreasing quadratic polynomial ($y=-2910(x - 0.0020)^2 + 5.27$) also fits very weakly ($R^2=0.0050$). For commercial files, there is an extremely weak negative correlation between the variables. A decreasing quadratic polynomial ($y=-6500(x - 0.0021)^2 + 6.59$) fits very weakly ($R^2=0.061$). 

For the combined set, there is a weak positive correlation between the physical center of vibrato and vibrato rate ($R=0.43,$ $R^2=0.18$). An increasing quadratic polynomial ($y=11.2(x - 6.94)^2 -536$) also fits weakly ($R^2=0.18$). Once again, due to the significant difference between the vibrato rate in commercial and TinySOL recordings, this combined fit is misleading.

Once again, the combined data set is not meaningful due to the discrepancy in vibrato rate between commercial and TinySOL recordings, as both data sets evenly span the spectrum of physical depths (Fig 5.). A Spearman's test for monotonicity for these data groups yields $\rho = -0.1780,$ and $p=0.1979$ for the commercial recordings, and $\rho = 0.0715,$ and $p=0.6612$ for the TinySOL recordings. As these $p$ values exceed any conventional threshold for statistical significance, we lack evidence to conclude a monotonic relationship between acoustical vibrato depth and vibrato rate.

Although acoustical vibrato depth might be expected to correlate negatively with vibrato rate\citep{TimmersDesain2000}—since the finger can only travel so far in a given time while preserving the integrity of the vibrato—these two results (Fig. S2, Fig. S3) suggest that performers rarely reach this physical limit.
\end{document}